\documentclass{blois}
\usepackage[utf8]{inputenc} 
\usepackage{siunitx}
\DeclareSIUnit\parsec{pc} 
\DeclareSIUnit\lightyear{ly}

\def\bbar{b\bar{b}}
\def\mumu{\mu^{+}\mu^{-}}
\def\tautau{\tau^{+}\tau^{-}}

\def\be{\begin{equation}}
\def\ee{\end{equation}}
\def\bea{\begin{eqnarray}}
\def\eea{\end{eqnarray}}
\newcommand{\ud}{\mathrm{d}}

\newcommand{\Photo}{}

\begin{document}
\vspace*{4cm}
\title{Dark Matter Searches with H.E.S.S.}

\author{ Knut Dundas Morå on behalf of the H.E.S.S. collaboration }

\address{ Stockholms universitet\\
Kosmologi astropartikelfysik och strängteori\\
106 91 Stockholm}

\maketitle\abstracts{
Cosmological measurements indicate that a large component of non-visible gravitating matter is present
in the universe. 
A common hypothesis for its origin is a weakly interacting, massive particle. Annihilations or decays of such particles
could produce gamma rays.
The H.E.S.S. experiment is an imaging air Cherenkov telescope array located in Namibia which may detect very high energy gamma-rays between 300 GeV and 10 TeV. 
This talk will present an overview of two recent H.E.S.S. searches
for dark matter in the very high energy region, one targeting dwarf galaxies, the other one a cored dark matter profile
at the galactic center. 
}

\section{Introduction}
Measurements of gravitational wells of galaxies and galaxy clusters cannot be explained by baryonic matter alone. Dark matter
that (almost) exclusively interacts through gravitational interactions, would explain both these and cosmological
observations well. Planck finds that the amount of dark matter is roughly 5 times that of baryonic matter~\cite{Ade:2015xua}.
If the dark matter is made up of particles that have weak interactions, they may be thermally produced in the early
universe with the observed abundance. For this reason, weakly interacting massive particles (WIMPs) are a widely studied
dark matter candidate. Two WIMPs may annihilate 
into standard model particles, which could lead to an observable
signal in gamma rays.

The High Energy Stereoscopic System (H.E.S.S.) is an array of four 12-meter 
and one 28-meter imaging Cherenkov telescopes located in the Namibian Khomas Highland.
In the two following analyses, only H.E.S.S. phase one data (four 12-meter telescopes) has been studied. 
High-energy cosmic rays and photons create showers in the atmosphere that emit Cherenkov light, which may be observed by fast cameras on the
telescopes. 
Details on the H.E.S.S. experiments may be found
in~\cite{refId0}. For typical analyses, the fiducial field of view is \ang{2} in diameter, and the effective area is 
$\sim\SI{1e5}{\meter\squared}$ above \SI{400}{\GeV}  . Under favorable conditions the analyses with H.E.S.S. I may be performed down to approximately \SI{300}{\GeV} 

This proceeding will summarize dark matter searches published by the H.E.S.S. collaboration the last year, one targeting
Dwarf Spheroidal Galaxies, the other the Galactic center.

\section{Search for Dark Matter in Dwarf Spheroidal Galaxies}
Dwarf Spheroidal Galaxies, dSphs, are satellites of the Milky Way. Inferred mass-to-light ratios up to several hundred from
stellar velocity measurements indicate that they are some of the most dark matter dominated systems in the universe.
Astrophysical backgrounds that might emit gamma rays are also low, making dSphs promising targets for indirect dark matter
searches. 
The H.E.S.S. experiment has searched for dark matter signals from five dSphs, with a total exposure of
$140$ hours ~\cite{Abramowski:2014tra} taken between 2006 and 2012. This is a fivefold increase with respect to the
previous search with dSphs published by H.E.S.S~\cite{2011APh....34..608H}. 
The Sagittarius dSph is the nearest of the five at \SI{25}{\kilo \parsec}, and was observed for \SI{90}{\hour}. 
The Coma Bernices, Fornax, Carina and Sculptor dSphs were observed with between $6$ and
\SI{12.7}{\hour} each.  

A circular signal region of $\theta \leq \ang{0.1}$ is chosen from the
instrument point spread function~\cite{refId0}. 
Background regions are constructed as a ring centered on the observation position~\footnote{see~\cite{refId0}}. 
Reconstruction of events, as well as gamma-hadron separation is done using the \emph{faint} selection of the
$X_{\mathrm{eff}}$ analysis~\cite{Dubois:2009zz}.

In order to compute limits on the dark matter cross section, the integrated dark matter content along the line-of-sight
$l$,  called the J-factor is needed: 
\begin{equation}
J =\frac{1}{\Delta \Omega}\int_l\int_{\Delta \Omega} \rho_{\mathrm{DM}}^2 \ud l \ud \Omega
\label{eq:jfac}
\end{equation}

In this analysis, J-factors and attendant uncertainties were using a novel Bayesian two-level likelihood that exploits
that dark matter parameters are shared by all dSphs to infer the dark matter content of individual dSphs. Details are
found in~\cite{2015MNRAS.451.2524M}.

The energy spectrum of the dark matter annihilation signal depends on what standard model particles take part. An
annihilation to two photons leads to a striking line feature, but this is usually loop-suppressed in comparison with
annihilations to leptons or quarks. In this analysis, the expected gamma ray spectra for annihilation into vector bosons, $\bbar$ and
leptons from~\cite{Cembranos:2010dm} are assumed.
A binned likelihood in energy for both OFF and ON spectra utilized the spectral information to improve the sensitivity. 

\begin{figure}[h]
\centerline{\includegraphics[width=0.53\linewidth]{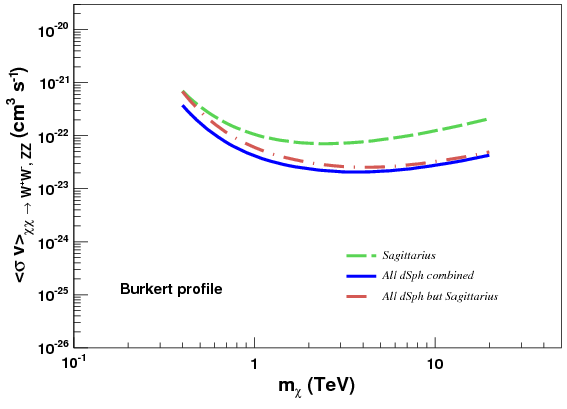}}
\caption[]{Limits on the dark matter self-annihilation cross section from five dwarf galaxies for annihilation into vector bosons.}
\label{fig:dwarfplot}
\end{figure}

No significant excess is observed between signal and background region for any of the five dSphs. The significances for
point sources across the fields of view are consistent with the background only hypothesis. Combined limits for all five dwarfs are set
using the profile likelihood procedure, and are shown in figure \ref{fig:dwarfplot}. The J-factor is included as a nuisance parameter, including the estimated
errors. The combined limit reaches $<\sigma v> \approx \SI{1.4e-23}{\centi\meter\cubed\per\second}$ for annihilation
into vector bosons. In the case of annihilation to $\mumu$ or $\tautau$ limits improve,
although not enough to reach a scan of supersymmetric models. 

\section{Search for Dark Matter from a Cored distribution in the Galactic Center}

\begin{figure}[h]
\centerline{\includegraphics[width=0.53\linewidth]{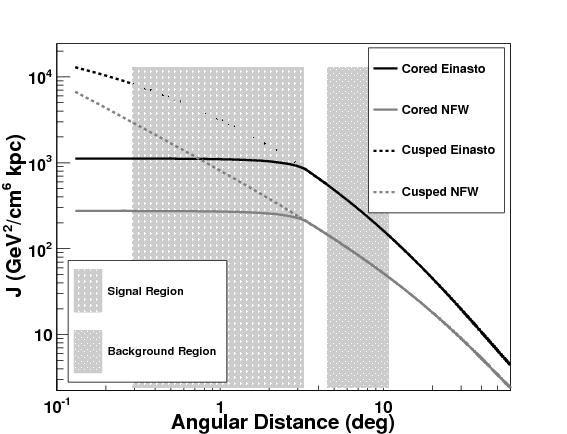}}
\caption{J-factors of the galactic center for cored and cusped dark matter profiles are displayed. The gray bands show the ON and OFF
regions.\label{fig:corejfactor}}
\end{figure}

Another search has been published using the Galactic center as the target~\cite{HESS:2015cda}.
N-body simulations of dark matter halos find that the density is sharply cusped, see fig.
\ref{fig:corejfactor}.
Previous analyses done with H.E.S.S. have searched for cusped
profiles~\cite{2011PhRvL.106p1301A} \cite{Abramowski:2013ax}.
However interactions with
baryonic matter could lead to a situation where the dark matter density is constant in the center of the
galaxy~\cite{Pontzen:2014lma}. 
Most observations done with H.E.S.S. use background and signal regions contained within the same
$\ang{2}\approx\SI{300}{\parsec}$ field of view.
A dark matter core could lead to roughly the same dark matter content in signal and background regions, reducing the
sensitivity. 
Therefore, this analysis uses separate background pointings, leading to the strongest limits on dark matter
annihilation without assuming a dark matter cusp. Limits are provided for the case of a core radius of 
\SI{500}{\parsec}, but can be translated to core radii up to \SI{2}{\kilo\parsec}.

In order to obtain OFF-regions outside a \SI{500}{\parsec} core, dedicated telescope pointings were done leading
to the signal and background coverage shown in figure \ref{fig:corejfactor}. Each observation of the ON-region was preceded and
followed by an OFF-run, equalizing the region of the atmosphere covered in each of the three runs. A total of
\SI{9}{\hour} of data were taken, \SI{3}{\hour} on the signal region.

\begin{figure}[h]
\centerline{\includegraphics[width=0.53\linewidth]{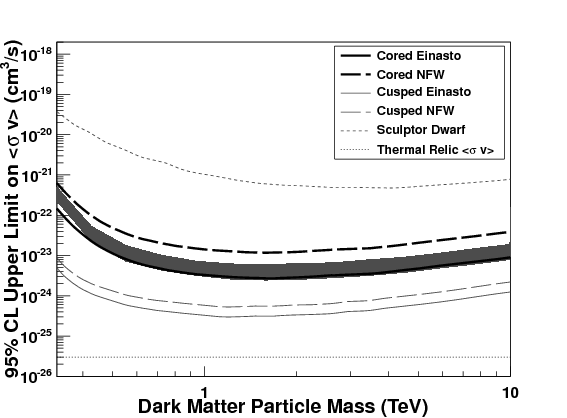}}
\caption{Upper limits on the annihilation cross-section of dark matter from the galactic center.\label{fig:corelimits}}
\label{fig:core}
\end{figure}

Events are selected and reconstructed as described in~\cite{refId0}.   
A cut in galactic latitude $|b| < \ang{0.3}$ is imposed to exclude astrophysical backgrounds in the Galactic plane. After
this cut, as well as image cleaning cuts, no excess is seen in sky maps of the signal or background positions.
The total excess of the signal region, including a $2\%$
uncertainty on the exposure ratio has a statistical significance of $-0.5\sigma$. 
As no significant signal is observed, upper limits on the velocity averaged dark matter
annihilation cross section are derived.  
Figure \ref{fig:corelimits} shows limits as function of dark matter mass. Cross sections above $<\sigma v> \approx
\SI{3e-24}{\centi\meter\cubed\per\second}$ are excluded in the mass region of highest sensitivity, between $1$ and
$\SI{4}{\tera\electronvolt}$, for a dark matter
density profile featuring a \SI{500}{\parsec} core.
If a cusped profile is assumed instead, the limits derived in this analysis improve by roughly a factor of 2. 

%
%

%

\section{Summary and Outlook}
The H.E.S.S. experiment has set limits on the annihilation of dark matter particles from the Galactic center and selected dwarf galaxies.
For the Galactic center, a dedicated pointing strategy let the experiment set the strongest current limit on the
annihilation of TeV-scale dark matter particles 
without the assumption of a cusped profile.
H.E.S.S.-II adds a single \SI{28}{\meter} telescope to the array. The larger dish area lowers the energy threshold of the
array, and will allow H.E.S.S.-II to bridge the gap between previous
searches and Fermi-LAT. 
The first
results of H.E.S.S.-II are being published, including preliminary work on updated dark matter
searches~\cite{Kieffer:2015nsa}~\cite{Lefranc:2015vza}.

\section*{References}


\begin{thebibliography}{10}

\bibitem{Ade:2015xua}
P.~A. R.~Ade et~al. (Planck~Collaboration).
\newblock {Planck 2015 results. XIII. Cosmological parameters}.
\newblock 2015.

\bibitem{refId0}
F.~Aharonian et~al. (H.E.S.S~Collaboration).
\newblock Observations of the {C}rab nebula with {H.E.S.S}.
\newblock {\em A\&A}, 457(3):899--915, 2006.

\bibitem{Abramowski:2014tra}
A.~Abramowski et~al. (H.E.S.S~Collaboration).
\newblock {Search for dark matter annihilation signatures in {H.E.S.S}
  observations of Dwarf Spheroidal Galaxies}.
\newblock {\em Phys. Rev.}, D90:112012, 2014.

\bibitem{2011APh....34..608H}
A.~Abramowski et~al. (H.E.S.S~Collaboration).
\newblock {{H.E.S.S} constraints on dark matter annihilations towards the
{S}culptor and {C}arina dwarf galaxies}.
\newblock {\em Astropart. Phys.}, 34:608--616, March 2011.

\bibitem{Dubois:2009zz}
F.~Dubois et~al.
\newblock {A multivariate analysis approach for the imaging atmospheric
  Cherenkov telescopes system {H.E.S.S}.}
\newblock {\em Astropart. Phys.}, 32:73--88, 2009.

\bibitem{2015MNRAS.451.2524M}
G.~D. Martinez.
\newblock {A robust determination of {M}ilky {W}ay satellite properties using
  hierarchical mass modelling}.
\newblock {\em MNRAS}, 451:2524--2535, August 2015.

\bibitem{Cembranos:2010dm}
J.~A. R.~Cembranos et~al.
\newblock {Photon spectra from WIMP annihilation}.
\newblock {\em Phys. Rev.}, D83:083507, 2011.

\bibitem{HESS:2015cda}
A.~Abramowski et~al. (H.E.S.S~Collaboration).
\newblock {Constraints on an Annihilation Signal from a Core of Constant Dark
  Matter Density around the Milky Way Center with {H.E.S.S}}
\newblock {\em Phys. Rev. Lett.}, 114(8):081301, 2015.

\bibitem{2011PhRvL.106p1301A}
A.~Abramowski et~al. (H.E.S.S~Collaboration).
\newblock {Search for a Dark Matter Annihilation Signal from the Galactic
  Center Halo with {H.E.S.S}}
\newblock {\em Phys. Rev. Lett.}, 106(16):161301, April 2011.

\bibitem{Abramowski:2013ax}
A.~Abramowski et~al. (H.E.S.S~Collaboration).
\newblock {Search for Photon-Linelike Signatures from Dark Matter Annihilations
  with {H.E.S.S}}
\newblock {\em Phys. Rev. Lett.}, 110:041301, 2013.

\bibitem{Pontzen:2014lma}
A.~Pontzen and F.~Governato.
\newblock {Cold dark matter heats up}.
\newblock {\em Nature}, 506:171--178, 2014.

\bibitem{Kieffer:2015nsa}
M.~Kieffer et~al. for~the H.E.S.S~Collaboration.
\newblock {Search for Gamma-ray Line Signatures with H.E.S.S}.
\newblock In {\em {Proceedings, 34th International Cosmic Ray Conference (ICRC
  2015)}}, 2015.

\bibitem{Lefranc:2015vza}
V.~Lefranc and E.~Moulin for~the H.E.S.S~Collaboration.
\newblock {Dark matter search in the inner Galactic halo with {H.E.S.S} I and
  {H.E.S.S} II}.
\newblock In {\em {Proceedings, 34th International Cosmic Ray Conference (ICRC
  2015)}}, 2015.

\end{thebibliography}

%
%
%
%


\end{document}